\documentclass[aps,pre,twocolumn,superscriptaddress,floatfix,secnumarabic]{revtex4-2}

\usepackage{graphicx}
\usepackage{amsmath,amssymb}
\usepackage{booktabs}
\usepackage{hyperref}

\begin{document}

\title{Collective search-and-capture under competing assignment policies}

\author{N\'estor Sep\'ulveda}
\affiliation{School of Engineering and Sciences, Universidad Adolfo Ib{\'a}{\~n}ez, Diagonal las Torres 2640, Pe\~{n}alolen, Santiago, Chile.}

\date{\today}

\begin{abstract}
We study a minimal lattice model of active search-and-capture in which
persistent random walkers locate and irreversibly capture immobile targets
through a finite-range, mutually exclusive assignment rule. We measure the
collective completion time $T_c$ as a function of the walkers'
reorientation rate $\alpha$ and the search radius $R$. The dependence
$T_c(\alpha)$ is non-monotonic, with a minimum at intermediate persistence
whose depth decreases as $R$ grows. Capture kinetics show that $T_c$ is not
a typical capture time but is governed by the extreme, late-time tail of
the capture process, while the bulk of targets are captured much earlier;
this tail is controlled mainly by the free-exploration phase rather than
by the final directed approach. We
then compare the baseline single-round assignment rule with cascading
reassignment and with maximum-cardinality matching on a candidate graph.
The two greedy policies (single-round and cascading) agree at very small
$R$, whereas maximum-cardinality matching already produces a strong
speedup at moderate $R$: improved matching reduces $T_c$ by factors of
several at large $R$, and by more than an order of magnitude at moderate
$R$. Thus, in this collective,
depletion-coupled search problem, the assignment policy can control the
capture time more strongly than the walkers' persistence.
\end{abstract}

\maketitle

\section{Introduction}

Populations of self-propelled or ``active'' particles exhibit collective
and single-particle behaviors with no equilibrium analogue, including
anomalous transport, motility-induced phase separation, and enhanced
search efficiency
\citep{Marchetti2013,Bechinger2016,Cates2015}. Much of this phenomenology
originates from persistent motion, whereby particles self-propel along a
body-fixed direction and only occasionally reorient, as first
characterized for the run-and-tumble dynamics of
\emph{Escherichia coli} \citep{Berg1972} and later formalized as a minimal
statistical-mechanics model \citep{Tailleur2008}. Minimal lattice models
have proved remarkably successful in revealing generic collective
phenomena generated solely by persistence, including nonequilibrium
wetting transitions, before incorporating additional ingredients relevant
to specific applications \citep{Sepulveda2017,Sepulveda2018}. More
recently, experiments have shown that surface geometry alone can strongly
reorganize the spatial distribution of persistent active particles,
leading to accumulation, depletion, or long-lived trapping
\citep{PerezEstay2024}.

Search and first-passage theory has established that persistence also
plays a central role in search efficiency. For a broad class of search
processes, an intermediate persistence length or reorientation rate
minimizes the mean search time
\citep{Redner2001,Benichou2011,Viswanathan1999,Tejedor2012}, although the
optimal strategy depends sensitively on the geometry of the environment
and on the spatial distribution of the targets \citep{Volpe2017}. These
results, however, almost exclusively concern the performance of
individual searchers or collections of independent searchers.

Here we ask how this familiar single-searcher optimum is modified when
many searchers and many targets interact through a finite-range,
mutually exclusive assignment rule and when targets disappear after
capture. The observable of interest is the time required to capture every
target, which is closely related to a cover time. Cover times for many
independent searchers have been characterized on networks
\citep{Kim2024}, with Gumbel-type limiting statistics for large target
sets \citep{Chupeau2015} and related results for the fastest of many
searchers \citep{Lawley2020}. Those theories, however, assume independent
walkers and do not include depletion, competition, or dynamic one-to-one
assignment. Our model adds precisely these ingredients through a matching
rule in the spirit of the classical assignment problem
\citep{Kuhn1955}.

We do not claim that the non-monotonic dependence of a search time on
reorientation is itself new. Instead, we determine how that optimum is
reshaped by target depletion and finite-range assignment, establish that
the completion time is governed by an extreme capture event, and quantify
how strongly alternative assignment policies modify the resulting
collective dynamics. Our central finding is that the assignment policy
reshapes this extreme-value statistic itself, not just its mean: policies
do not primarily accelerate the already-directed final approach to a
target; instead, they suppress the long intervals during which the last
few unmatched walkers and targets fail to establish a productive
assignment, which is what the completion time is ultimately governed by.
Beyond the minimal search-and-capture setting studied here, this is the
same qualitative trade-off, between local and centrally coordinated
matching, faced by foraging or collecting agents, search-and-rescue
robots, and dispatch systems that assign mobile agents to spatially
distributed demand.

\section{Model}

\paragraph{Lattice and populations.}
The system is a periodic square lattice of $L_x\times L_y=40\times40$
sites. At $t=0$, $N_w=\phi L_xL_y$ mobile walkers and
$N_t=\phi_gL_xL_y$ immobile targets are placed with
$\phi=\phi_g=0.3$, so that $N_w=N_t=480$. Targets occupy distinct sites.
Walkers are placed independently by rejection sampling on sites not
occupied by targets; multiple walkers may share a site.

\paragraph{Free dynamics.}
Each walker carries an orientation
$v\in\{\mathrm{N},\mathrm{S},\mathrm{E},\mathrm{W}\}$. During each time
step, walkers move one lattice spacing in their current direction, in a
randomly reshuffled order, with periodic wrap-around. After all walkers
have moved, each orientation is independently redrawn uniformly from the
four possible directions with probability $\alpha$ (and so left unchanged,
with probability $1/4$, even on a redraw) and otherwise remains
unchanged. In the absence of targets, this is a discrete-time persistent
random walk with mean redraw interval $1/\alpha$; because a redraw
conserves the current direction one time in four, the mean interval
between actual direction changes is $4/(3\alpha)$.

\paragraph{Capture and internal state.}
Every walker carries a binary state $s\in\{0,1\}$, initialized at $s=0$.
When a walker with $s=0$ lands on a live target, the target is removed and
the walker irreversibly switches to $s=1$. A walker with $s=1$ continues
to move but no longer interacts with targets. Since $N_w=N_t$ and each
walker captures at most one target, complete capture is possible.

\paragraph{Baseline assignment rule.}
Before movement, every live target identifies its nearest walker with
$s=0$ within Euclidean distance $R$, using the periodic minimum-image
convention. If several targets select the same walker, the walker keeps
only its closest target; rejected targets are not reassigned during that
step. The resulting single-round assignment is therefore mutually
exclusive but generally non-maximal. Each assigned walker is steered for
that step along the cardinal direction that most reduces its distance to
the target; ties between the two axes are broken uniformly at random. The
assignment is recomputed from scratch every step. Complete algorithmic
details, tie-breaking rules, and the alternative policies are given in the
Supplemental Material.

\paragraph{Completion time and protocol.}
Let $N_{\mathrm{live}}(t)$ denote the number of targets not yet captured
at time $t$. We define the completion time as
\begin{equation}
T_c=\min\{t : N_{\mathrm{live}}(t)=0\}.
\end{equation}
The baseline scan covers
$\alpha\in\{0.01,0.02,\ldots,0.50\}$ and $R\in\{1,2,\ldots,30\}$, with
$100$ independent realizations per parameter pair, for up to
$1.5\times10^5$ simulations. Every realization converged before a safety
cutoff of $2\times10^6$ time steps.

\section{Results}

\subsection{Persistence and assignment range}

\begin{figure*}[htbp]
  \centering
  \includegraphics[width=\textwidth]{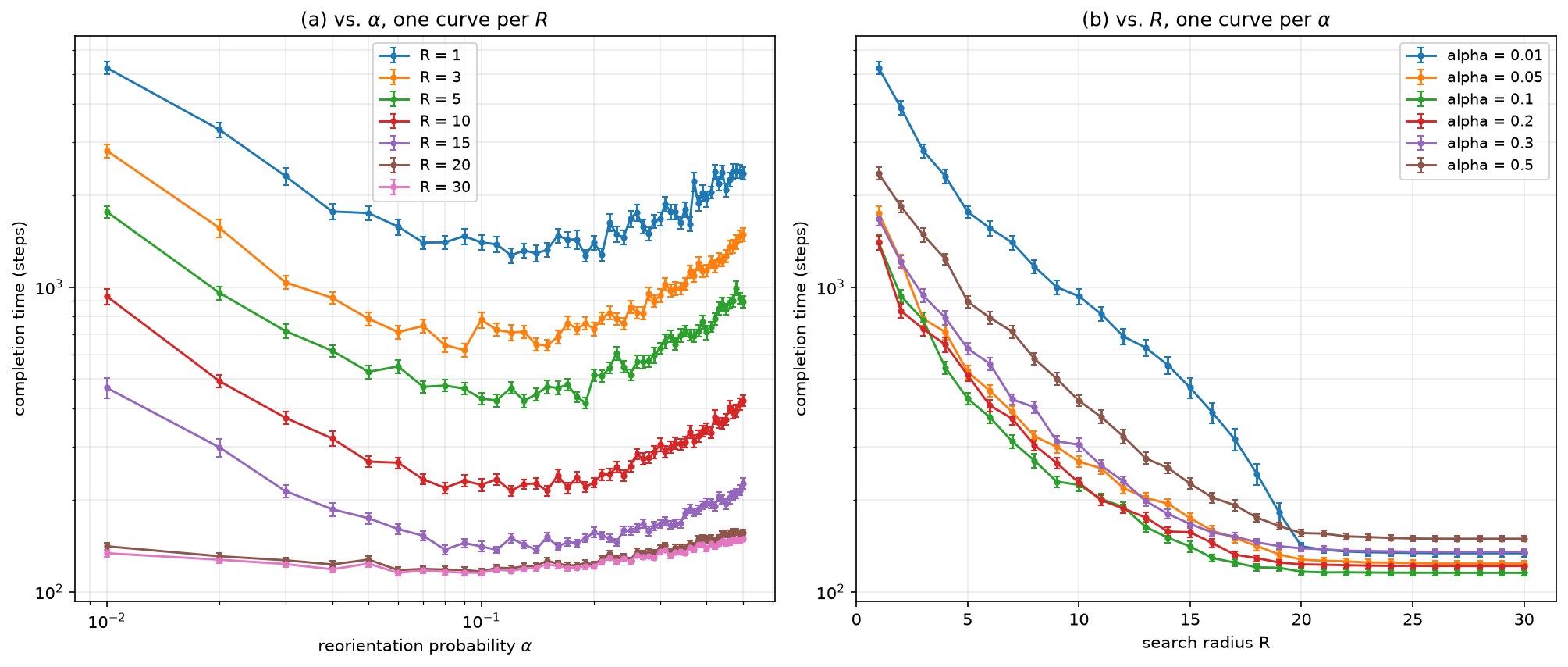}
  \caption{Completion time over a representative subset of the full
  $50\times30$ $(\alpha,R)$ grid (mean $\pm$ SEM over $100$ replicas;
  $L=40$; the complete grid, with all sampled $R$ and $\alpha$ and with
  s.d.\ error bars, is given in the Supplemental Material). (a)
  $T_c$ versus reorientation probability $\alpha$, one curve per
  search radius $R$. (b) $T_c$ versus $R$, one curve per $\alpha$.
  The curves flatten as $R$ approaches the largest possible periodic
  minimum-image separation, $L/\sqrt{2}\approx28$ for $L=40$.}
  \label{fig:main-results}
\end{figure*}

Figure~\ref{fig:main-results}(a) shows a pronounced intermediate-$\alpha$
minimum of $T_c(\alpha)$ at small and intermediate $R$; the minimum
becomes shallow at large $R$. At $R=1$, the grid minimum is
near $\alpha^\ast\approx0.19$, with
$T_c=1269\pm572$ (mean $\pm$ s.d.\ across replicas, not SEM), compared
with $5248\pm2497$ at $\alpha=0.01$ and
$2360\pm1226$ at $\alpha=0.5$. At $R=10$, the minimum lies near
$\alpha^\ast\approx0.12$, where $T_c=214\pm80$. As $R$ increases, the
minimum becomes much shallower: for $R\gtrsim20$, changing $\alpha$ across
the entire sampled range changes $T_c$ by only about $30$--$40\%$.

The exact grid location of $\alpha^\ast$ is less robust than the existence
of the minimum itself. Bootstrap confidence intervals are broad whenever
the curve is shallow, whereas the overall flattening of $T_c(\alpha)$ with
increasing $R$ is statistically clear. Details of the bootstrap analysis,
the window-dependent power-law fits, and the full table of minima are
reported in the Supplemental Material.

Figure~\ref{fig:main-results}(b) shows that $T_c$ decreases with $R$ and
crosses over to a finite-size plateau. On a periodic square lattice, the
largest possible minimum-image separation is $L/\sqrt2$, so increasing $R$
beyond this geometric scale cannot expose additional walker--target pairs.
The crossover begins before the hard ceiling and scales with $L$, as shown
in the Supplemental Material.

\subsection{The completion time is an extreme capture time}
\label{sec:kinetics-main}

For each target, let $T_i$ denote its capture time. The global completion
time can be written as
\begin{equation}
T_c=\max_{i=1,\ldots,N_t} T_i.
\end{equation}
We further decompose each capture time as
\begin{equation}
T_i=t_i^{\mathrm{start}}+\tau_i^{\mathrm{steer}},
\end{equation}
where $t_i^{\mathrm{start}}$ is the step at which the walker that
captures target $i$ begins the continuous, uninterrupted assignment run
that ends in that capture (any earlier assignment to a different target,
or an intervening step spent unassigned, resets this clock), and
$\tau_i^{\mathrm{steer}}=T_i-t_i^{\mathrm{start}}$ is the duration of that
final directed approach; $t_i^{\mathrm{start}}$ is therefore tied to the
target actually captured, not to whichever target this walker was first
ever assigned to.

\begin{table}[htbp]
\centering
\caption{Capture kinetics at the $\alpha$ values used in the dedicated
kinetics campaign ($100$ replicas; $N_t=480$), chosen near the minima of
the baseline scan (Table~S1 of the Supplemental
Material) but not always identical to its grid-optimal $\alpha^\ast$,
since that campaign was run separately and independently rounded. The
latest assignment-start time in the population closely tracks $T_c$,
whereas $T_{90\%}\ll T_c$.}
\label{tab:kinetics-main}
\begin{tabular}{@{}ccccccc@{}}
\toprule
$R$ & $\alpha$ & $T_{50\%}$ & $T_{90\%}$ & $T_c$ & $\max(t^{\rm start})$ & $\max(t^{\rm start})/T_c$ \\
\midrule
1  & 0.15 & 0.4 & 18.2 & 1324 & 1324 & 1.000 \\
5  & 0.12 & 0.4 & 9.6  & 467  & 462  & 0.990 \\
10 & 0.12 & 0.4 & 9.5  & 214  & 203  & 0.947 \\
20 & 0.08 & 0.4 & 9.4  & 118  & 98   & 0.828 \\
30 & 0.08 & 0.4 & 9.4  & 116  & 94   & 0.808 \\
\bottomrule
\end{tabular}
\end{table}

We index time so that $t=0$ is the first opportunity for capture,
immediately after the first move; no capture is possible before it, since
no walker starts on an occupied target site. $T_{50\%}$ and $T_{90\%}$
(Table~\ref{tab:kinetics-main}) are, per replica, the integer step at
which the $50$th and $90$th percentile of captures occur, averaged over
$100$ replicas, so a value like $T_{50\%}=0.4$ is a replica-averaged step
count, not a fractional time within one run: the fraction of targets
captured at $t=0$ itself is $F_1\approx0.50$ at every $R$ tested,
essentially $R$-independent because it is set by pairs that already start
at distance $1$. $T_{90\%}$ stays small ($9$--$18$ steps) across the
table, while $T_c$ is one to two orders of magnitude larger. Thus, $T_c$
carries little information about a typical capture and is instead
governed by the extreme, late-time tail of the process, qualitatively
consistent with the Gumbel-type extreme-value statistics found for many
independent searchers \citep{Chupeau2015,Kim2024}, though we have not
tested that specific distributional form directly, and the present
walkers are correlated through depletion and competitive assignment,
unlike the independent-searcher setting those results assume.

The latest assignment-start time in the population accounts for most of
$T_c$ (ratio $1.00$ at $R=1$, falling to $0.81$ at $R=30$; Table
\ref{tab:kinetics-main}), whereas the mean steering time is of order one lattice step, remaining
below $1.1$ steps at every $R$ tested. The instantaneous
fraction of searching walkers that are assigned to a target provides a
second, population-level test of the same mechanism.

\begin{figure}[htbp]
  \centering
  \includegraphics[width=\linewidth]{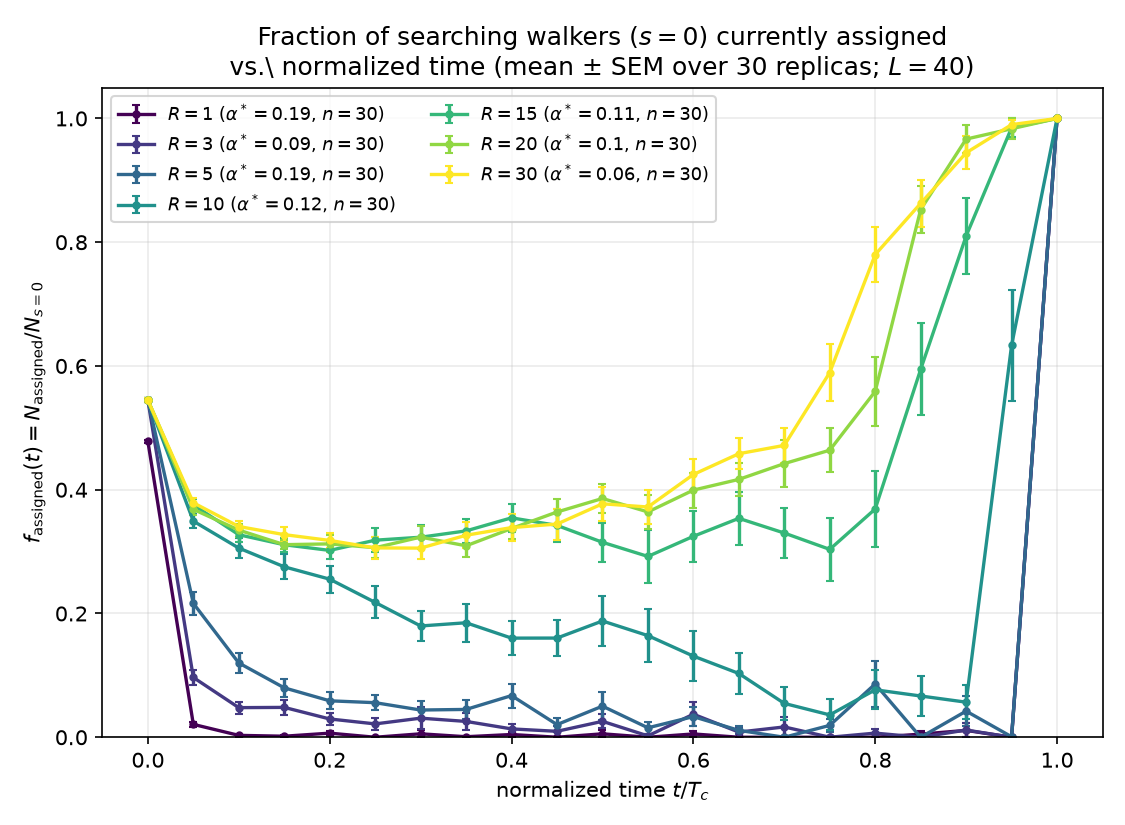}
  \caption{Fraction of searching walkers currently assigned to a target,
  $f_{\mathrm{assigned}}(t)$, versus normalized time $t/T_c$ (mean $\pm$
  SEM over $30$ replicas; each realization is linearly interpolated onto a
  common $t/T_c$ grid before averaging). At small $R$, most remaining
  walkers spend nearly the whole process in free exploration; at large
  $R$, assignment remains substantial throughout the dynamics. The final
  rise toward unity is partly an artifact of conditioning on $T_c$ itself
  and of the small, shrinking number of still-searching walkers as
  $t\to T_c$ (Sec.~S5.1 of the Supplemental
  Material), and should not be read as a sharp collective transition.}
  \label{fig:assignfrac-main}
\end{figure}

Figure~\ref{fig:assignfrac-main} shows that at every $R$, the baseline
rule assigns approximately half of the searching walkers already at
$t=0$, $f_{\rm assigned}(0)\approx0.5$. This is a related but distinct
observation from the $F_1\approx0.50$ fraction of targets already
captured by $t=0$ (Sec.~\ref{sec:kinetics-main}, an assigned walker only
captures immediately if its target is at distance exactly $1$, and a
minority of captures happen without any active assignment): both are set
predominantly by walker--target pairs already adjacent at $t=0$, which
any $R\geq1$ both detects and, when adjacent, captures in a single move,
so neither fraction depends strongly on $R$. At $R=1$, the assigned fraction then
rapidly falls below $0.05$ and remains small for most of the process. At
$R=30$, it stays of order $0.3$--$0.4$ through the middle of the run and
rises earlier toward unity, though the endpoint itself should be read
with the caveat above. These results
show that increasing $R$ progressively replaces the slow, fluctuation-
dominated free-exploration phase with directed motion.

\subsection{Assignment policies control the collective capture time}
\label{sec:policy-main}

We compared the baseline single-round policy with two alternatives.
Cascading reassignment sorts walker--target pairs within $R$ by distance
and greedily accepts pairs whose walker and target are both still free.
The third policy computes a maximum-cardinality matching on a candidate
graph using augmenting paths \citep{Kuhn1955}. For computational
tractability, candidate lists are truncated at large $R$; it is therefore
an exact maximum matching only on the resulting candidate graph. Full
algorithmic details and controls are provided in the Supplemental
Material.

\begin{figure}[htbp]
  \centering
  \includegraphics[width=\linewidth]{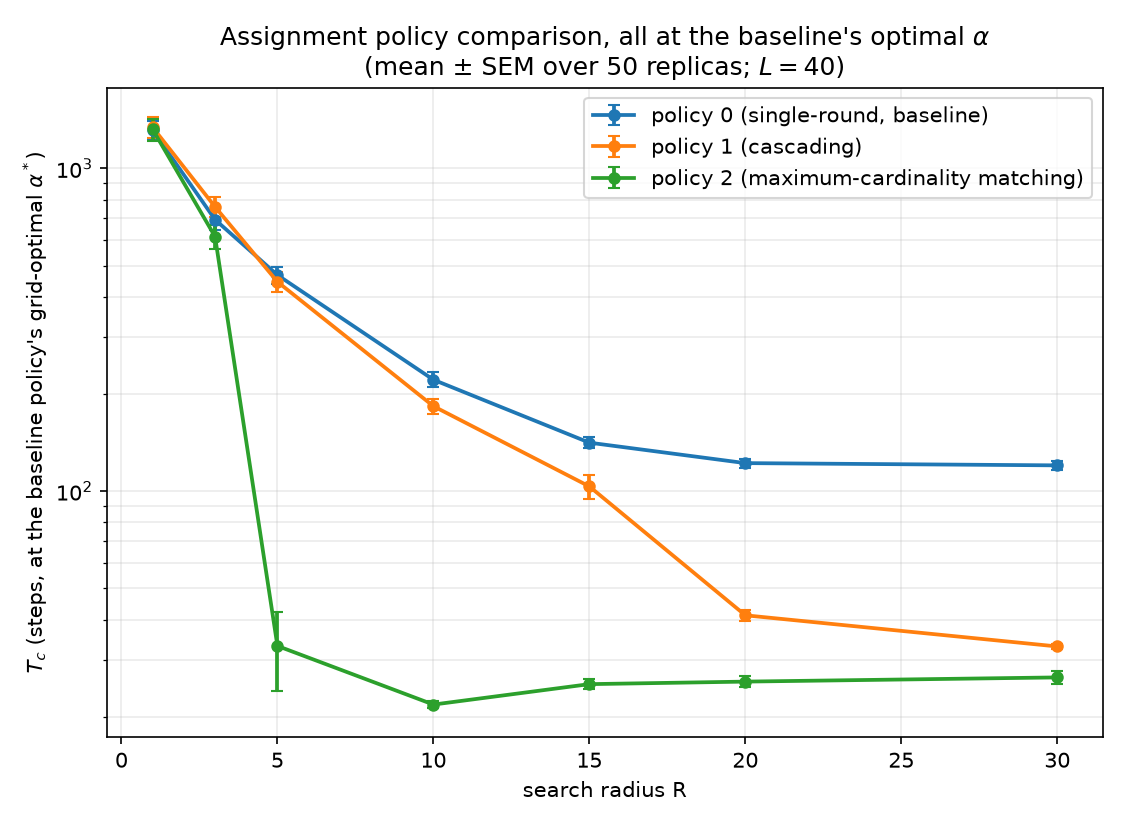}
  \caption{Near-optimal completion time for the baseline single-round,
  cascading, and maximum-cardinality assignment policies (mean $\pm$ SEM
  over $50$ replicas; $L=40$). The two greedy policies (single-round and
  cascading) agree at very small $R$; maximum-cardinality matching
  diverges from both already at moderate $R$.}
  \label{fig:policy-main}
\end{figure}

Figure~\ref{fig:policy-main} shows that at $R\leq5$, the baseline and
cascading policies agree within uncertainty.
From $R\approx10$ onward, cascading reassignment becomes significantly
faster, reducing $T_c$ to roughly one third of the baseline value by
$R=20$--$30$. This effect persists when each policy is evaluated at its
own optimal $\alpha$ and across larger lattices. The Supplemental
Material gives the corresponding speedup factor,
$T_c^{\mathrm{baseline}}/T_c^{\mathrm{policy}}$, directly against $R$ for
both alternative policies.

Maximum-cardinality matching produces an even stronger reduction. At
$R=5$, the completion time falls from approximately $467$ steps for the
baseline and $445$ for cascading reassignment to about $33$ steps. The
assignment policy therefore changes $T_c$ by more than the entire
persistence optimum does at the same $R$. The mechanism is contention:
when many targets share candidate walkers, a single-round policy leaves
many targets unassigned, whereas a more complete matching uses the
available walkers more efficiently and suppresses the long
free-exploration tail. A dedicated instrumented campaign, reported in the
Supplemental Material, confirms directly (not merely from the expected
candidate count) that the candidate graph is untruncated at $R=5$, and
shows that maximum-cardinality matching reassigns walkers to different
targets roughly $30$ times more often, per step, than either heuristic.
The speedup therefore cannot be explained solely by the larger initial
assignment fraction: maximum-cardinality matching also sustains a far
more dynamically reorganized assignment throughout the run, which is
associated with (though not, by this correlation alone, proven to cause)
the suppression of the late free-exploration tail. Because the
augmenting-path algorithm selects one specific matching among possibly
several of the same maximum cardinality, we also verified that
randomizing its internal tie-breaking order leaves this conclusion
intact: the speedup over the baseline remains a full order of magnitude
($9$--$14\times$ depending on tie-breaking, always excluding no
improvement by a wide margin), so it is not an artifact of a particular
degenerate matching (Supplemental Material).

\section{Discussion}

The non-monotonic dependence of $T_c$ on $\alpha$ is consistent with the
usual trade-off between ballistic and diffusive search. Walkers that
reorient too rarely remain confined to long straight runs, while walkers
that reorient too often lose the extended excursions needed to explore
new regions efficiently. In the present model, however, this familiar
single-particle effect is embedded in a many-body completion time that is
governed by the last target and strongly modified by the assignment rule.

The decomposition
$T_i=t_i^{\mathrm{start}}+\tau_i^{\mathrm{steer}}$ provides a useful
effective picture. The measured steering contribution is short, while the
latest assignment-start time in the population closely follows $T_c$.
Increasing $R$, adding redundant walkers, or improving the matching
policy all act primarily by reducing the free-exploration-dominated tail.
The Supplemental Material shows that the
saturation radius scales with the lattice size, that no statistically
resolved dependence of $\alpha^\ast$ on $L$ remains after bootstrap
analysis, and that varying the walker-to-target ratio produces the same
qualitative flattening of the persistence optimum.

A simple independent-target extreme-value estimate,
$T_c\sim(\log N_t)/\lambda_{\mathrm{eff}}$, does not describe the
finite-size data quantitatively. This indicates that the effective capture
rate itself varies with system size, as expected when target depletion and
competitive assignment correlate the late capture events. A complete
theory of this correlated extreme-value process remains open.

\section{Conclusion}

We introduced a many-searcher, many-target lattice model combining
persistent exploration, irreversible capture, and dynamic one-to-one
assignment. The collective completion time has an intermediate-persistence
minimum, but this minimum becomes shallow as the search radius grows.
More importantly, $T_c$ is an extreme capture time dominated by the final
target, and the tail is controlled mainly by free exploration rather than
the directed final approach.

Alternative assignment policies strongly reshape this tail. Cascading
reassignment reduces the completion time by factors of several at large
$R$, while maximum-cardinality matching can reduce it by more than an order
of magnitude at moderate $R$. The three policies also differ in the
coordination they require: single-round is local and requires no
communication beyond a target's immediate neighborhood, cascading needs a
one-shot global sort of candidate pairs, and maximum-cardinality matching
requires solving a global combinatorial problem over the full candidate
graph every step. In this model, access to greater coordination in
solving the assignment problem can therefore be a stronger control
parameter for collective capture speed than optimizing the walkers'
persistence.


\bibliographystyle{apsrev4-2}
\bibliography{references}

\end{document}